\documentclass[useAMS,usenatbib]{mn2e}
\usepackage{multirow}
\usepackage{epsfig}
\usepackage{subfigure}

\makeatletter
\let\jnl@style=\rm
\def\ref@jnl#1{{\jnl@style#1}}

\def\aj{\ref@jnl{AJ}}                   
\def\araa{\ref@jnl{ARA\&A}}             
\def\apj{\ref@jnl{ApJ}}                 
\def\apjl{\ref@jnl{ApJ}}                
\def\apjs{\ref@jnl{ApJS}}               
\def\ao{\ref@jnl{Appl.~Opt.}}           
\def\apss{\ref@jnl{Ap\&SS}}             
\def\aap{\ref@jnl{A\&A}}                
\def\aapr{\ref@jnl{A\&A~Rev.}}          
\def\aaps{\ref@jnl{A\&AS}}              
\def\azh{\ref@jnl{AZh}}                 
\def\baas{\ref@jnl{BAAS}}               
\def\jrasc{\ref@jnl{JRASC}}             
\def\memras{\ref@jnl{MmRAS}}            
\def\mnras{\ref@jnl{MNRAS}}             
\def\pra{\ref@jnl{Phys.~Rev.~A}}        
\def\prb{\ref@jnl{Phys.~Rev.~B}}        
\def\prc{\ref@jnl{Phys.~Rev.~C}}        
\def\prd{\ref@jnl{Phys.~Rev.~D}}        
\def\pre{\ref@jnl{Phys.~Rev.~E}}        
\def\prl{\ref@jnl{Phys.~Rev.~Lett.}}    
\def\pasp{\ref@jnl{PASP}}               
\def\pasj{\ref@jnl{PASJ}}               
\def\qjras{\ref@jnl{QJRAS}}             
\def\skytel{\ref@jnl{S\&T}}             
\def\solphys{\ref@jnl{Sol.~Phys.}}      
\def\sovast{\ref@jnl{Soviet~Ast.}}      
\def\ssr{\ref@jnl{Space~Sci.~Rev.}}     
\def\zap{\ref@jnl{ZAp}}                 
\def\nat{\ref@jnl{Nature}}              
\def\iaucirc{\ref@jnl{IAU~Circ.}}       
\def\aplett{\ref@jnl{Astrophys.~Lett.}} 
\def\apspr{\ref@jnl{Astrophys.~Space~Phys.~Res.}}
\def\bain{\ref@jnl{Bull.~Astron.~Inst.~Netherlands}}
\def\fcp{\ref@jnl{Fund.~Cosmic~Phys.}}  
\def\gca{\ref@jnl{Geochim.~Cosmochim.~Acta}}   
\def\grl{\ref@jnl{Geophys.~Res.~Lett.}} 
\def\jcp{\ref@jnl{J.~Chem.~Phys.}}      
\def\jgr{\ref@jnl{J.~Geophys.~Res.}}    
\def\jqsrt{\ref@jnl{J.~Quant.~Spec.~Radiat.~Transf.}}
\def\memsai{\ref@jnl{Mem.~Soc.~Astron.~Italiana}}
\def\nphysa{\ref@jnl{Nucl.~Phys.~A}}   
\def\physrep{\ref@jnl{Phys.~Rep.}}   
\def\physscr{\ref@jnl{Phys.~Scr}}   
\def\planss{\ref@jnl{Planet.~Space~Sci.}}   
\def\procspie{\ref@jnl{Proc.~SPIE}}   

\makeatother

\title[The NGC~3341 minor merger]{The NGC~3341 minor merger: a panchromatic view of the active galactic nucleus in a dwarf companion}

\author[Stefano Bianchi, et al.]{Stefano Bianchi$^1$\thanks{E-mail: bianchi@fis.uniroma3.it (SB)}, Enrico Piconcelli$^{2}$, Miguel \'Angel P\'erez-Torres$^3$, 
\newauthor Fabrizio Fiore$^2$, Fabio La Franca$^1$, Smita Mathur$^{4,5}$, Giorgio Matt$^1$\\
$^1$Dipartimento di Matematica e Fisica, Universit\`a degli Studi Roma Tre, via della Vasca Navale 84, 00146 Roma, Italy\\
$^2$Osservatorio Astronomico di Roma - INAF, Via di Frascati 33, 00040, Monte Porzio Catone, RM, Italy\\
$^3$IAA-CSIC, Glorieta de la Astronom\'ia s/n, 18008 Granada, Spain\\
$^4$Department of Astronomy, The Ohio State University, Columbus, Ohio 43210, USA\\
$^5$Center for Cosmology and Astro-Particle Physics, The Ohio State University, Columbus, OH 43210\\
}

\begin{document}

\maketitle

\label{firstpage}

\begin{abstract}
We present X-ray (\textit{Chandra}), radio (\textit{EVLA} and \textit{EVN}), and archival optical data of the triple-merging system in NGC~3341. Our panchromatic analysis confirms the presence of a Seyfert 2 AGN in NGC~3341B, one of the secondary dwarf companions. On the other hand, the nucleus of the primary galaxy, consistent with a star-forming region of a few solar masses per year, and NGC~3341C are very unlikely to host an AGN. We therefore suggest that NGC~3341 is an exceptional example of an AGN triggered in the satellite galaxy of a minor-merging system. The existence of such a system can have important implications in the models of hierarchical growth of structures. Further observational and theoretical efforts on NGC~3341 and potentially similar sources are needed in order to understand the role of minor mergers on the onset of AGN activity, and in the evolution of massive galaxies.
\end{abstract}


\section{Introduction}

Supermassive black holes (SMBHs) are found at the heart of most nearby  galaxies. Dynamical evidence indicates that their masses tightly correlate with the stellar velocity dispersion of the host bulge \citep[the so--called $M_{\rm BH}$--$\sigma_*$ relation;][]{fermer00} revealing a fundamental connection between SMBHs and galaxy evolution:  growth of SMBHs and build--up of galaxies are closely related \citep{dimatteo05,mh08}. Multiple Mergers offer a potential physical mechanism linking galaxy star formation, morphology, and AGN (i.e. active SMBHs) evolution \citep[e.g.][]{hopkins08}. Mergers and interactions can induce large-scale radial motion of the interstellar gas within the  host galaxy, and trigger its infall into the innermost central regions to fuel the AGN and to enhance the star-formation activity in the merger remnants \citep[e.g.][]{ellison11}.

``Seed'' SMBHs would keep on growing by further mergers and gas accretion, evolving into the population of bright AGN observed at lower redshifts ($z \sim$ 2--4) and, finally, into the ubiquitous SMBH relics among massive galaxies in the local Universe. If galaxies merge hierarchically and all galactic bulges contain SMBHs, the formation of  binary/multiple SMBHs should be therefore inevitable \citep{hq04}. Interestingly, the presence of binary SMBHs has been also linked to many important aspects of the AGN phenomenon \citep[e.g.][for a review]{kom03}, i.e. the formation of the molecular torus (key ingredient of the Unified models), the difference between radio-loud and radio-quiet AGN, the distortions and the bendings in radio jets, and the random orientations of the radio jets and bi-conical narrow-line regions with respect to the rotational axes of the host galaxy disk. Finally, coalescing binary SMBHs are expected to be the most powerful sources of gravitational waves.

AGN trace the active, easily-observable phase of SMBH and, therefore, they clearly are the ideal object where discovering SMBH. The term ``binary AGN'' indicates a system in which two active nuclei form a gravitationally linked pair at orbital separation of $\leq$10 pc  while a galaxy with two widely separated active nuclei is called a ``dual AGN''.
The longest timescale of the binary evolution is represented by a phase where the nuclei are at a separation of 0.1--1 pc. Unfortunately, such objects cannot be spatially resolved with the present telescopes \citep[e.g.][]{mm05} for a review). On the other hand, any observational evidence even for dual AGN is so far very rare. Most of the known candidates have large separations (tens of kpc), and were found thanks to the SDSS \citep[e.g.][]{djorg87,koch99,hennawi06,myers08,green10}. On the other hand, the high penetrative power of hard X--ray observations provides a unique and often ultimate tool in the hunt for multiple active nuclei in a galaxy. The first clear-cut example in the X-rays was found in the Ultra-luminous Infrared Galaxy (ULIRG) NGC~6240 \citep{kom03b}, followed by a handful of other confirmed cases \citep{ballo04,gua05c,bianchi08c,evans08,pico10}, including the very close pair (only 150 pc) found recently in NGC~3393 \citep{fabb11}. A common property of these dual AGN is that they are heavily dust-enshrouded nuclei. In addition some members of these systems show no (or very weak) explicit AGN evidence in their optical/near-IR spectra. 

NGC~3341 is a nearby ($z$=0.027) disturbed system consisting of a minor merger between a giant disk galaxy ($M_{\rm B}$=$-$20.3 mag) and two low-mass dwarf companions within the galaxy disk (see left panel of Fig.~\ref{3341_field}), with a stellar mass ratio of 50:2:1 \citep[primary:B:C, ][B08 hereafter]{barth08}. The two offset companions are at projected distance of 5.1 kpc (9.5 arcsec) (NGC~3341B; $M_{\rm B}$=$-$17 mag) and 8.4 kpc (15.6 arcsec) (NGC~3341C; $M_{\rm B}$=$-$16.6 mag) from the nucleus and may be dwarf ellipticals or the spheroidal merger remnant of low-mass spirals. The separation between B and C is $\sim$12 arcsec i.e. $d$=6.4 kpc. On the basis of Sloan Digital Sky Survey observations and follow-up Keck II spectroscopy, B08 confirmed the physical association between the three objects (based on the common radial velocity) and discovered the surprising presence of a Seyfert 2 nucleus in the dwarf satellite NGC~3341B. On the other hand, according to the $\mathrm{[{O\,\textsc{iii}}]}$/H$\beta$ vs. $\mathrm{[{N\,\textsc{ii}}]}$/H$\alpha$ emission-flux diagnostic diagram, the primary  nucleus can be classified as a composite LINER/$\mathrm{{H\,\textsc{ii}}}$ object. Its nature is, however, still uncertain as it is consistent with an $\mathrm{{H\,\textsc{ii}}}$ nucleus on the basis of another diagnostic diagram. Finally, NGC~3341C show emission line-ratios typical of LINERs but it was not possible to definitively assess the presence of an active SMBH at its center.  

NGC~3341 had never observed so far in the X-rays, with the exception of a non-detection in the \textit{ROSAT} All-Sky survey, which corresponds to a loose upper limit of $\mathrm{F_{0.1-2\,keV}}\simeq3\times10^{-13}$ erg cm$^{-2}$ s$^{-1}$ (B08), which is largely compatible with the expected X-ray flux for this class of object. In this paper, we present X-ray (\textit{Chandra}), radio (\textit{EVLA} and \textit{EVN}), and archival optical data of this very peculiar system, in order to unveil the nature of its components.

\section{Observations and data reduction}

In the following, errors and upper limits correspond to the 90 per cent confidence level for one interesting parameter, where not otherwise stated. The adopted cosmological parameters are $H_0=70$ km s$^{-1}$ Mpc$^{-1}$ , $\Omega_\Lambda=0.73$ and $\Omega_m=0.27$ \citep[i.e. the default ones in \textsc{xspec 12.7.1}:][]{xspec}.

\subsection{Chandra}

NGC~3341 was observed by \textit{Chandra} on 2012, November 9, for a total exposure time of 50 ks (ObsID 13871), with the Advanced CCD Imaging Spectrometer \citep[ACIS;][]{acis}. Data were reduced with the Chandra Interactive Analysis of Observations \citep[CIAO;][]{ciao} 4.5 and the Chandra Calibration Data Base (CALDB) 4.5.5 software, adopting standard procedures. Spectra and count-rates were extracted from circular regions of 2 arcsec of radius, apart from NGC~3341B (2.5 arcsec), the background evaluated from circular regions of 5 arcsec of radius. No source has enough counts for a rebinning which allows us to use the $\chi^2$ statistics. We therefore opted for the \citet{cash76} statistics in the following fits to X-ray data.

\subsection{SDSS}

The  Sloan Digital Sky Survey (SDSS-III) Data Release 9 (DR9) provided us with images in different filters (u: 3551 \AA, g: 4686 \AA, r: 6166 \AA, i: 7480 \AA, z: 8932 \AA) of the NGC3341 field, and an optical spectrum of NGC3341B (Plate: 0578, MJD: 52339, Fiber: 0495), with a wavelength coverage of 3800-9200 \AA, and a resolution of 1500 at 3800 \AA. Further details can be found at the SDSS-III website\footnote{http://www.sdss3.org}.

\subsection{VLA}

We used publicly available, unpublished radio observations of NGC~3341 taken with the Expanded Very Large Array (EVLA) on August and September 2009 (ID code AU126) at a frequency of 4.86 GHz. Since all observations were taken while the EVLA was  in C-configuration, we only present here the results of the deepest observations, which  were carried out on 2009 August 14, when the total on-target time was approximately 4.5 hr.
 The absolute flux density calibrator was 1331+305, and the source 1041+061 was used as phase-calibrator. We used standard calibration procedures within the NRAO software package \emph{AIPS}. We also used {\it AIPS} for imaging purposes, using standard hybrid mapping techniques.
We reached an off-source r.m.s. figure of 15 $\mu$Jy/b, using a natural weighting scheme.
However, our final image was obtained using a uniform weighting scheme, which yielded a finer synthesized beam (full-width at half maximum of 4.2$\times$3.8 arcsec along a position angle of -58 deg) at the expense of a higher r.m.s. (26 $\mu$Jy/beam). This allowed us to marginally disentangle the radio emission in the vicinity of the Nucleus, NGC~3341B and NGC~3341C, which otherwise appears tangled in the naturally-weighted image.

\subsection{EVN}

We observed NGC~3341 on 2013 March 19th and 2013 April 16th with the electronic European VLBI Network (eEVN) at the frequencies of 5.0 GHz (code RSP08; PI: P\'erez-Torres) and 1.7 GHz (code RSP08B; PI: P\'erez-Torres), respectively. At  5.0 GHz, the array included the following nine antennas: Effelsberg (Germany), Jodrell Bank (Mk~2; UK), Noto (Italy), Onsala (Sweden), Torun (Poland), Yebes (Spain), Westerbork (The Netherlands), and Hartebeestoek (South Africa). At 1.7 GHz, the array was the following: Effelsberg, Jodrell Bank (Lovell; UK), Medicina (Italy), Onsala, Torun, Westerbork (The Netherlands), and Hartebeestoek. 

The observations were taken at each frequency in the most sensitive mode currently sustained by the eEVN, i.e., 1 Gigabit/sec and 2-bit sampling. The antennas observed in dual polarization, using 16 bands of 16 MHz each, which in turn were split into 32 spectral channels of 0.5 MHz each, for a total synthesized bandwidth of 256 MHz, and the integration time was of 2 sec. These technicalities allowed us to image the several-arcsecond field around NGC 3341 with milliarcsecond resolution, while minimizing the effects of bandwidth- or time-smearing.

We observed in phase-referenced mode, alternatively observing the strong ($\sim$400 mJy at 5.0 GHz, and $\sim$760 mJy at 1.6 GHz), nearby phase-calibrator J1038+512 and our target source, NGC~3341, with a typical duty cycle of 5 min. To calibrate the band-pass, we used OQ~208 at 5.0 GHz, while at 1.7 GHz we used our phase-calibrator, J1038+512, which was strong enough for band-pass calibration purposes.
We used standard calibration and imaging procedures within the \emph{AIPS} package of NRAO, including flagging, amplitude and parallactic angle correction, deriving phase, delay and phase-rate solutions, band-pass calibration, and hybrid-mapping of the phase-calibrator and our target source, NGC~3341. We also corrected for the electron content of the ionosphere, which at our observing frequencies may have a significant effect in the measured phases of the sources. 

At 5.0 GHz, our eEVN observations resulted in a synthesized beam of (11.0 mas $\times$ 6.0 mas) at a position angle of 47 degrees. The final r.m.s. global noise figure in the image  was of 21 $\mu$Jy/b, although there were some regions where the r.m.s figure was larger, likely due to the poor coverage of the interferometric ($u-v$) plane and other calibration issues which could not be fully removed from the data. In particular, in the region around the nucleus, we found a source with flux density 81$\mu$Jy/b ($\sim$3.9 the global r.m.s.) coincident within a few milliarcseconds with the position of the X-ray nucleus (and that of the radio position) of NGC~3341. The local r.m.s around NGC~3341B and NGC~3341C was 26 $\mu$Jy/b, and we did not find any source brighter than three times the r.m.s. and close to their X-ray counterparts. 

At 1.7 GHz, the synthesized beam was of (16.7 mas $\times$ 13.3 mas) at a position angle of -61 degrees, and the final r.m.s. was $\sim$70 $\mu$Jy/b around the regions of the nucleus, NGC~3341B, and of NGC~3341C. This noise figure was significantly larger than expected, mainly due to the corrupted data taken at the Jodrell Bank (Lovell) and Hartebeestoek telescopes, as well as due to significant radio frequency interference (RFI) at many antennas, which severely decreased the sensitivity and fidelity of the final image. We found no evidence of emission above 3$\times$r.m.s. in the vicinity of the nucleus of NGC~3341, neither around NGC~3341B, nor NGC~3341C. We note, however, that given the large r.m.s. figure at 1.6 GHz, we cannot exclude that there is an underlying radio emitting component. 

\section{The NGC3341 field}

\begin{figure*}
\begin{center}
\epsfig{file=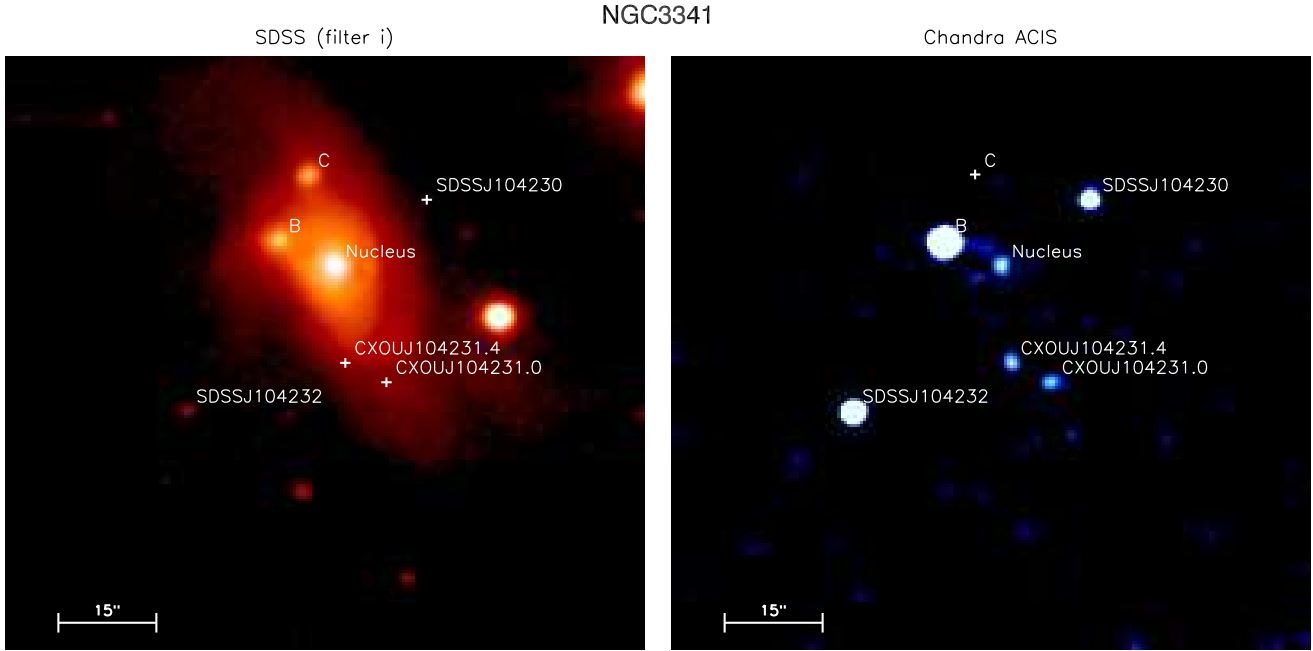,width=\textwidth}
\end{center}
\caption{\label{3341_field}NGC~3341: SDSS (filter i: \textit{left}) and \textit{Chandra} (0.2-8 keV: \textit{right}) images with labels as in Table~\ref{Chandrasources}. Sources undetected in either image are also marked by a white cross.}
\end{figure*}

\begin{table*}
\caption{\label{Chandrasources}The sources detected (apart from the upper limit on source C) by \textit{Chandra} in the NGC~3341 field shown in Fig.~\ref{3341_field}. See text for details.}
\begin{center}
\begin{tabular}{lccccccc} \hline
Name & RA & DEC & z& $CR_s$ & $CR_h$ & $F_t$ & $L_t$\\
(1) & (2) & (3) & (4) & (5) & (6) & (7) & (8) \\
\hline
Nucleus & 10:42:31.47 & +05:02:37.63 & 0.027339 & $4.1^{+1.7}_{-1.3}$ & $0.6^{+0.8}_{-0.5}$ & $2.5^{+1.0}_{-0.8}$ & $0.43^{+0.17}_{-0.14}$\\
NGC3341B & 10:42:32.05 & +05:02:41.51 & 0.027154 & $3.8^{+1.6}_{-1.4}$ & $110\pm8$ & $270\pm19$ & $46\pm3$\\
NGC3341C & 10:42:31.75 & +05:02:52.83 & 0.027169 & $<0.7$ & $<0.5$ & $<0.18$ & $<0.03$\\
SDSS J104232.99+050213.7 & 10:42:33.00 & +05:02:13.47 & 1.675 & $22\pm3$ & $8\pm2$ & $20\pm3$ &  $4.0\pm0.6\times10^4$\\
SDSS J104230.54+050248.6 & 10:42:30.57 & +05:02:48.89 & -- & $4.2^{+3.7}_{-1.7}$ & $3.6^{+1.6}_{-1.3}$ & $6.6\pm1.7$ & --\\
CXOU J104231.4+050221 & 10:42:31.37 & +05:02:21.86 & -- & $1.2^{+1.1}_{-0.8}$ & $2.4^{+1.3}_{-1.0}$ & $4.8^{+2.2}_{-1.8}$ & -- \\
CXOU J104231.0+050218 & 10:42:30.97 & +05:02:18.67 & -- & $1.8^{+1.3}_{-0.9}$ & $1.1^{+1.0}_{-0.7}$ & $1.9^{+1.1}_{-0.9}$ & --\\
\hline
\end{tabular}
\end{center}

\textsc{Notes.}-- Col. (1) Name of the source - (2), (3) RA and DEC (J2000) from \textit{Chandra} if detected, from \textit{SDSS} (filter i) otherwise - (4) Redshift (from NED, apart from NGC3341C: B08) (5), (6) Soft (0.5-2 keV) and hard (2-7 keV) count-rates in units of $10^{-4}$ c/s - (7) Total (0.5-7 keV) flux in units of $10^{-15}$ erg cm$^{-2}$ s$^{-1}$ (8) Total (0.5-7 keV) luminosity (not corrected for absorption) in units of $10^{40}$ erg s$^{-1}$
\end{table*}

\begin{figure}
\begin{center}
\epsfig{file=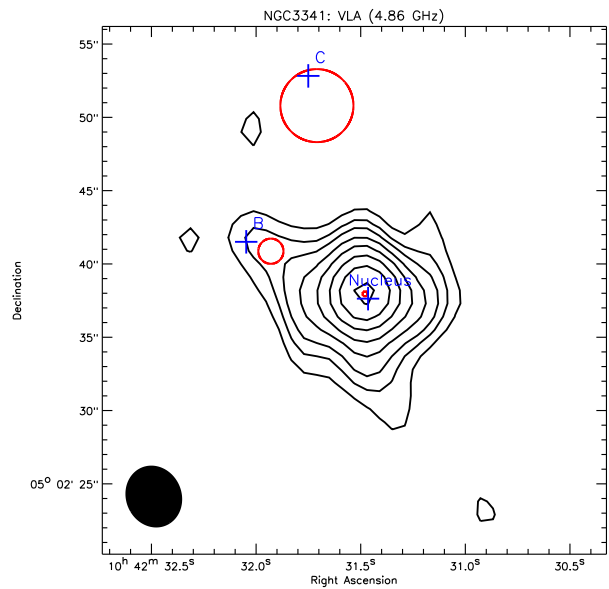,width=0.45\textwidth}
\end{center}
\caption{\label{3341_vla}NGC~3341: EVLA observation (4.86 GHz), with the beam size indicated in the lower left corner. Blue crosses indicate the \textit{Chandra} positions (\textit{SDSS} filter i for source C), red circles the VLA detections (both at 99\% astrometric accuracy). The VLA detection for source C is marginal ($<3\sigma$).}
\end{figure}

\begin{table*}
\caption{\label{evla_log}\textit{EVLA} radio observations of the NGC~3341 field (see Fig.~\ref{3341_vla}). Upper limits correspond to three times the r.m.s. in the 5.0 GHz \textit{EVLA} image.}
\begin{center}

\begin{tabular}{lrrrrr} \hline
Name & Peak flux & Total flux  &  $L_{\nu, peak}$ & $L_{\nu, int}$& $\nu\, L_{\nu, \rm int}$ \\
(1)      & (2)           & (3)            & (4)                                  & (5)                            & (6) \\
\hline
Nucleus      &  1048     & 2380        &  17.5 $\pm 0.6$& 39.7$\pm$0.9 & 19.7$\pm$0.5\\
NGC3341B &    172     &   259        &    2.9 $\pm$0.4 & 4.3$\pm$0.4   &  1.4$\pm$0.2 \\
NGC3341C &$\leq$78 &  $\leq 78$&    $\leq$1.3 & ---  & $\leq$0.6\\
\hline
\end{tabular}
\end{center}

\textsc{Notes.}-- Col. (1) Name of the source, as in Table 1
- (2) Peak flux density, in $\mu$Jy/beam
- (3) Total flux density, in $\mu$Jy obtained from a single Gaussian fit
- (4) 5.0 GHz Peak monochromatic radio luminosity, in units of $10^{27}$ erg s$^{-1}$ Hz$^{-1}$
- (5) 5.0 GHz Integrated monochromatic radio luminosity, in units of $10^{27}$ erg s$^{-1}$ Hz$^{-1}$
- (6) Total 5.0 GHz radio luminosity, in units of $10^{37}$ erg s$^{-1}$
\end{table*}

\subsection{\label{3341analysis}The nucleus, NGC3341B, and NGC3341C}

The nucleus of NGC3341 is clearly detected by \textit{Chandra}, even if only marginally in the hard (2-7 keV) X-ray band (see right panel of Fig.~\ref{3341_field} and Table~\ref{Chandrasources}). However, the number of counts is too limited to allow for a meaningful spectral analysis. We therefore computed a hardness ratio $\mathrm{HR}\equiv\frac{H-S}{H+S}$, where $H$ and $S$ are the net counts in the hard (2-7 keV) and soft (0.5-2 keV) band, respectively. We adopted a Bayesian estimation of HR and its uncertainties as in \citet{park06}. The resulting value of HR$=-0.76^{+0.12}_{-0.16}$ is consistent with a very steep ($\Gamma>2.2$) spectrum. Indeed, most of the X-ray emission is emitted below 2 keV, for a total (0.5-7 keV) luminosity of $\simeq4\times10^{39}$ erg s$^{-1}$. On the other hand, the nucleus is by far the brightest source in the radio image provided by \textit{EVLA} (see Fig.~\ref{3341_vla}). Its total 5 GHz luminosity is $\simeq2\times10^{38}$ erg s$^{-1}$ (see Table~\ref{evla_log}). However, the 5.0 GHz \textit{EVLA} flux is completely resolved out by our 5.0 GHz \textit{EVN} observations, and there is no sign of any compact source, except for a very faint ($\mathrm{L_{5\,GHz}}\simeq6.6\times10^{36}$ erg s$^{-1}$) unresolved radio source at the 3.9-sigma level. Our 1.7 GHz \textit{EVN} observations did not show evidence for any compact source, either.

\begin{figure}
\begin{center}
\epsfig{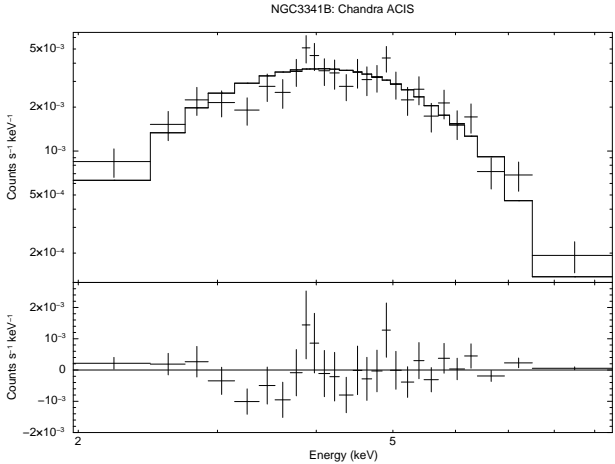}
\end{center}
\caption{\label{3341B_Chandra}NGC~3341B: \textit{Chandra} ACIS spectrum with best fit model and residuals. See text for details.}
\end{figure}

\begin{figure}
\begin{center}
\epsfig{file=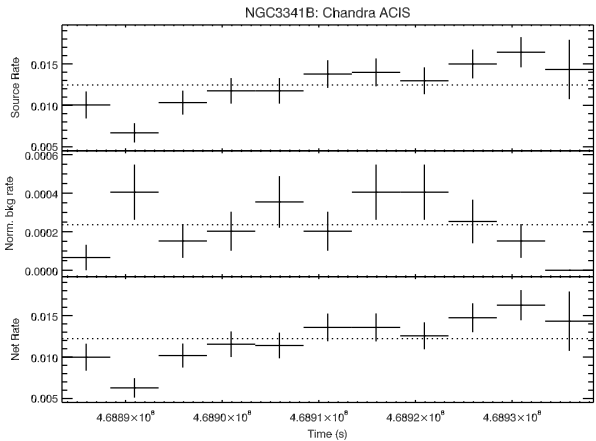,width=\columnwidth}
\end{center}
\caption{\label{3341B_lc}NGC~3341B: \textit{Chandra} ACIS lightcurve (0.2-8 keV) from the source extraction region (\textit{top}), the background extraction region normalized to the latter (\textit{middle}), and the background-subtracted source extraction region (\textit{bottom}). }
\end{figure}

NGC3341B is the X-ray brightest object in the field (see right panel of Fig.~\ref{3341_field} and Table~\ref{Chandrasources}). Its \textit{Chandra} spectrum can be well fitted by a simple power-law, fixed at $\Gamma=1.7$ \citep[e.g.][]{bianchi09} because it cannot be constrained in the fit, absorbed by a column density $N_H=1.21^{+0.14}_{-0.13}\times10^{23}$ cm$^{-2}$ at the redshift of the source (see Fig.~\ref{3341B_Chandra}). The 2-10 keV flux of this model is $5.0\pm0.4\times10^{-13}$ erg cm$^{-2}$ s$^{-1}$, corresponding to an absorption-corrected luminosity of $1.54\pm0.12\times10^{42}$ erg s$^{-1}$. No neutral iron K$\alpha$ line is detected, with an upper limit of $2.5\times10^{-6}$ ph cm$^{-2}$ s$^{-1}$ to its flux (EW$<250$ eV). A very faint soft X-ray emission is detected below the photoelectric cut-off at $\simeq2$ keV, consistent with the non-detection by \textit{ROSAT} (B08). During the \textit{Chandra} observation, NGC3341B displayed some variability (see Fig.~\ref{3341B_lc}).

\begin{figure*}
\begin{center}
\epsfig{file=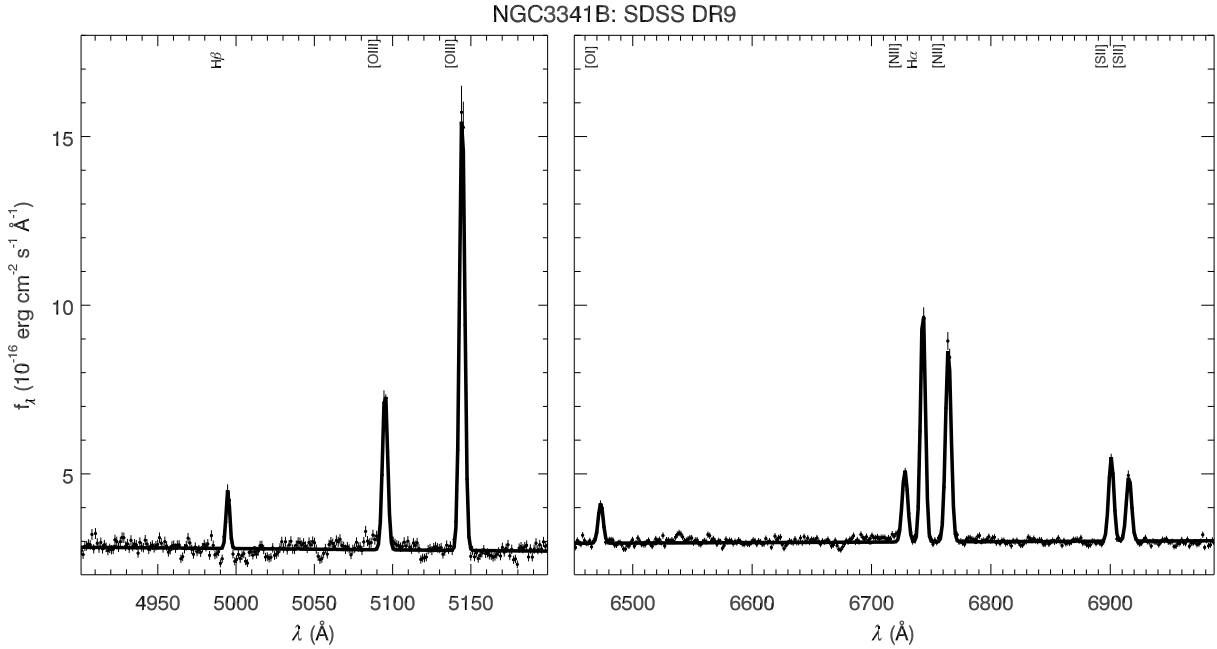,width=\textwidth}
\end{center}
\caption{\label{3341B_sdss}NGC~3341B: the SDSS optical spectrum, with the best fit emission lines reported in Table~\ref{ngc3341B_sdss}.}
\end{figure*}

\begin{figure*}
\begin{center}
\epsfig{file=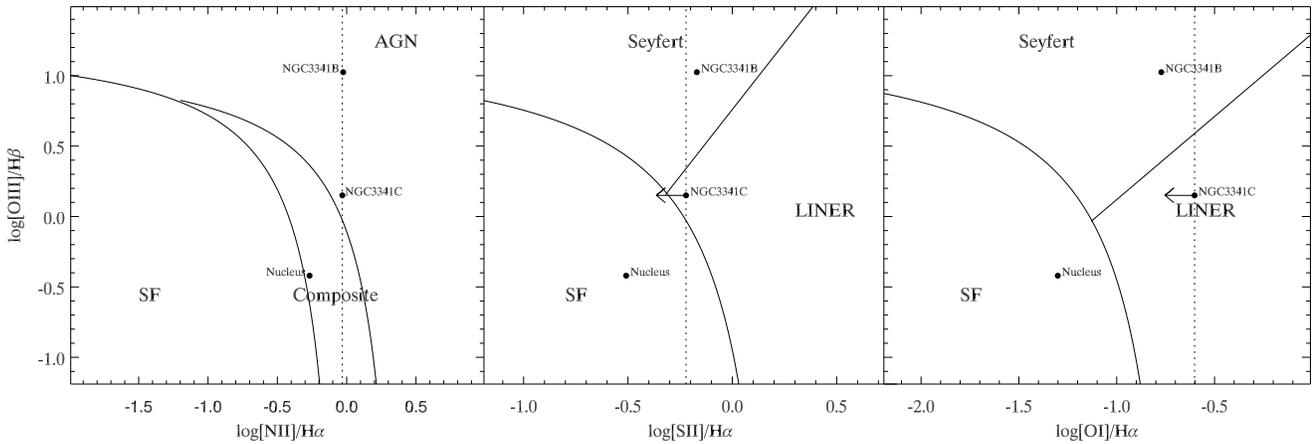,width=\textwidth}
\end{center}
\caption{\label{3341B_bpt}Optical line diagnostic diagrams for the nucleus of NGC~3341 and NGC~3341C (data taken from B08), and NGC~3341B (this work). Classifications are done after \citet{kew06}, and references therein. The dotted line for NGC~3341C is due to the unconstrained value for the $\mathrm{[{O\,\textsc{iii}}]}/\mathrm{H}\beta$ ratio, the $\mathrm{[{O\,\textsc{iii}}]}$ emission line being undetected in B08.}
\end{figure*}

The \textit{SDSS} optical spectrum is shown in Fig.~\ref{3341B_sdss}, and the detected emission lines listed in Table~\ref{ngc3341B_sdss}. No broad components of the Balmer lines are present. Standard diagnostics based on lines ratio unequivocally indicate that NGC3341B is a Seyfert 2 galaxy (Fig.~\ref{3341B_bpt}), in agreement with the results reported by B08 on the basis of their Keck spectrum. The Balmer decrement H$\alpha/$H$\beta\simeq7$ leads to a de-reddened $\mathrm{[{O\,\textsc{iii}}]}$ luminosity of $\simeq1.1\times10^{41}$ erg s$^{-1}$, thus $\log \left(L_x/L_{\mathrm{[{O\,\textsc{iii}}]}}\right)\simeq1.1$, in perfect agreement with the values generally found in Compton-thin Seyfert galaxies \citep[see e.g.][]{lamastra09}. Adopting bolometric corrections appropriate for X-ray \citep{mar04} and $\mathrm{[{O\,\textsc{iii}}]}$ \citep{lamastra09} luminosities, we get a consistent bolometric luminosity for NGC3341B of $\simeq2\times10^{43}$ erg s$^{-1}$. With an estimated BH mass of $2\times10^6$ M$\odot$ (from stellar velocity dispersion: B08), this results in a $L_{bol}/L_{Edd}\simeq0.08$.  

\begin{table}
\caption{\label{ngc3341B_sdss}NGC3341B: optical emission lines in the \textit{SDSS} spectrum (see Fig.~\ref{3341B_sdss})}
\begin{center}
\begin{tabular}{llll}
Line & $\lambda_\mathrm{l}$ & FWHM & Flux \\
(1) & (2) & (3) & (4)\\
\hline
 &  &  &\\
H$\beta$ & 4861.33 & $160\pm30$ & $5.1\pm0.7$ \\[1ex]
$\mathrm{[{O\,\textsc{iii}}]}$ & 4958.92 & $220\pm10$ & $19.0\pm1.5$\\[1ex]
$\mathrm{[{O\,\textsc{iii}}]}$ &  5006.85 & $220\pm10$ & $54\pm3$ \\[1ex]
$\mathrm{[{O\,\textsc{i}}]}$ & 6300.32 & $210\pm30$ & $6.0\pm0.5 $\\[1ex]
$\mathrm{[{N\,\textsc{ii}}]}$ & 6548.06 & $230\pm10$ & $12.0\pm1.0$ \\[1ex]
H$\alpha$ & 6562.79 & $210\pm10$ & $35.5\pm1.5$  \\[1ex]
$\mathrm{[{N\,\textsc{ii}}]}$ & 6583.39 & $230\pm10$ & $33.5\pm1.5$ \\[1ex]
$\mathrm{[{S\,\textsc{ii}}]}$ & 6716.42 & $215\pm10$ & $13.5\pm1.0$\\[1ex]
$\mathrm{[{S\,\textsc{ii}}]}$ & 6730.78 & $215\pm10$ & $10.5\pm1.0$\\
 &  &  &  \\
\hline
\end{tabular}
\end{center}

\textsc{Notes.}-- Col. (1) Identification. Col (2) Laboratory wavelength (\AA),
in air \citep{bowen60}. Col. (3) km s$^{-1}$ (instrumental resolution not removed).
 Col. (4) $10^{-16}$ erg cm$^{-2}$ s$^{-1}$.
\end{table}

NGC3341B is also detected in the radio \textit{EVLA} image, although not completely resolved from the brighter nucleus (see Fig.~\ref{3341_vla} and Table~\ref{evla_log}), and therefore might have some contribution from the overall nuclear region. Assuming all of the detected 4.86 GHz radio emission is intrinsic to NGC 3341B, its 4.86 GHz luminosity is $\simeq1.4\times10^{37}$ erg s$^{-1}$. This value  is well consistent with its X-ray luminosity, when compared to the observed correlation in bright Seyfert galaxies \citep[e.g.][]{bianchi09b,lmf10}, although somewhat fainter than expected when compared to the same correlation in local AGN with lower luminosity \citep[e.g.][]{panessa07}. Any unresolved radio source at the spatial resolution of the EVN is too faint to be detected down to a 5.0 GHz luminosity of $\simeq6.1\times10^{36}$ erg s$^{-1}$ (3 times the r.m.s.), which is not uncommon in nearby Seyfert galaxies \citep{pg13}.

NGC3341C is not detected neither by \textit{Chandra}, with an upper limit to its total (0.5-7 keV) luminosity of $\simeq3\times10^{38}$ erg s$^{-1}$ (see Table~\ref{Chandrasources}), nor by the \textit{eVLA}, with an upper limit ($3\sigma$) to its 5.0 GHz luminosity of $6\times10^{36}$ erg s$^{-1}$ (see Table~\ref{evla_log}). 

\begin{figure}
\begin{center}
\epsfig{file=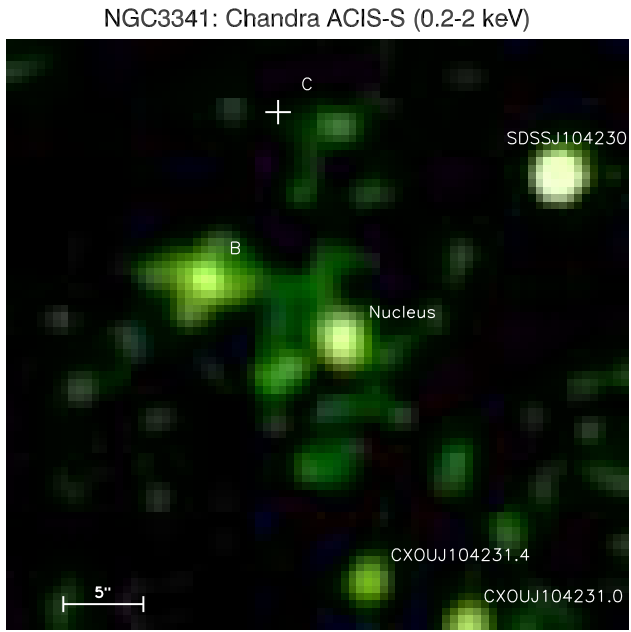,width=\columnwidth}
\end{center}
\caption{\label{3341_softX}The \textit{Chandra} 0.2-2 keV image of NGC~3341: extended emission around NGC~3341B and the nucleus is observed, together with a faint emission connecting the two sources. The \textit{SDSS} (filter i) position of source C is marked by a cross.}
\end{figure}

Finally, in Fig.~\ref{3341_softX} a \textit{Chandra} soft X-ray (0.2-2 keV) image of the system is shown. It is clear that extended emission is present around NGC~3341B. This is in agreement with what commonly found in Seyfert 2 galaxies, and interpreted as gas photoionised by the AGN, roughly coincident with the optical NLR \citep[e.g.][]{kin02,yws01,bianchi06,gb07,ghosh07}. Soft X-ray extended emission is also present south-west of the nucleus, consistent with star formation as expected from the optical and radio emission properties of this source, together with fainter emission connecting the latter to NGC~3341B, which may be also related to shocked gas as in the merger systems NGC~6240 \citep{fer13} and NGC~7771+NGC~7770 \citep{alherr12}. Deeper high-resolution X-ray observations are needed to investigate the origin of this emission, possibly coupled with spatially resolved optical spectroscopy in $\mathrm{[{O\,\textsc{iii}}]}$ and H$\alpha$.

\subsection{The other sources in the NGC~3341 field}

\begin{figure}
\begin{center}
\epsfig{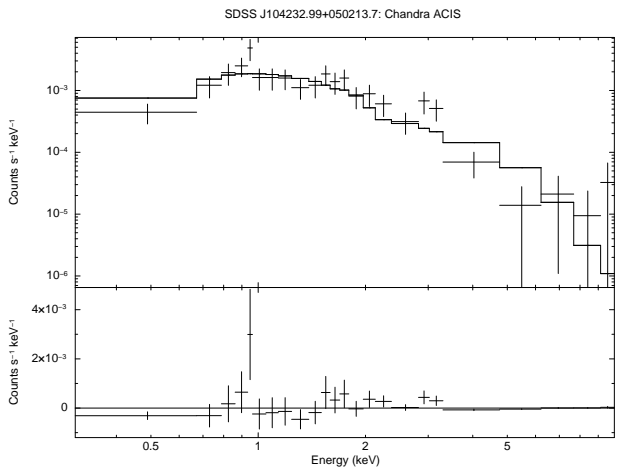}
\end{center}
\caption{\label{s1_Chandra}SDSS J104232.99+050213.7: \textit{Chandra} ACIS spectrum with best fit model and residuals. See text for details.}
\end{figure}

\begin{figure}
\begin{center}
\epsfig{file=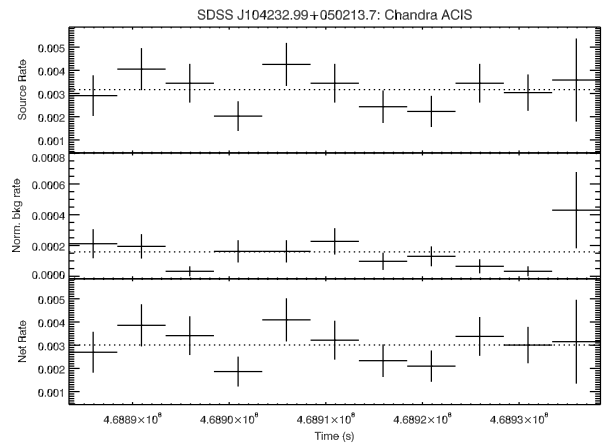,width=\columnwidth}
\end{center}
\caption{\label{s1_lc}SDSS J104232.99+050213.7: \textit{Chandra} ACIS lightcurve (0.2-8 keV) from the source extraction region (\textit{top}), the background extraction region normalized to the latter (\textit{middle}), and the background-subtracted source extraction region (\textit{bottom}).}
\end{figure}

The \textit{Chandra} spectrum of SDSS J104232.99+050213.7, shown in Fig.~\ref{s1_Chandra} can be fitted by a power-law ($\Gamma=1.7^{+0.3}_{-0.2}$), absorbed by the Galactic column density \citep[$2.56\times10^{20}$ cm$^{-2}$:][]{lab}. This model has a 2-10 keV flux of $1.9\pm0.7\times10^{-14}$ erg cm$^{-2}$ s$^{-1}$. No neutral iron K$\alpha$ line is detected, with an upper limit of $1.1\times10^{-6}$ ph cm$^{-2}$ s$^{-1}$ to its flux (EW$<450$ eV). This source was classified as a quasar candidate by \citet{rich04}, with a photometric redshift of 1.675. Considering also the reported uncertainties  on the redshift, the absorption-corrected 2-10 keV luminosity is therefore $1.6\pm0.9\times10^{44}$ erg s$^{-1}$. No significant variability is observed during the \textit{Chandra} observation (Fig.~\ref{s1_lc}).

Although SDSS J104230.54+050248.6 is detected by \textit{Chandra} both in the soft and in the hard X-ray bands, the total number of counts is very low, and no meaningful information can be extracted from a spectral analysis. The hardness ratio HR$=-0.26^{+0.14}_{-0.15}$  is consistent with a somewhat flat ($\Gamma\simeq1-1.6$) or mildly absorbed (up to $6\times10^{21}$ cm$^{-2}$ if $\Gamma=1.8$) spectrum.
Its 0.5-7 keV flux is $\simeq7\times10^{-15}$ erg cm$^{-2}$ s$^{-1}$ (see Table~\ref{Chandrasources}). There is no further information on this source in the literature, apart from the SDSS magnitudes:  23.4 (u), 25.1 (g), 24.9 (r), 22.6 (i), 21.6 (z).

CXOU J104231.4+050221 and CXOU J104231.0+050218 are weaker sources, with no counterpart in the SDSS, nor in the literature. Their hardness ratios HR are $0.34^{+0.22}_{-0.27}$ and $-0.23^{+0.22}_{-0.31}$, respectively. They are consistent with an extremely flat ($\Gamma<0.6$) or rather absorbed (few $10^{22}$ cm$^{-2}$ if $\Gamma=1.8$) spectrum for the first source, and an unabsorbed $\Gamma\simeq1.8$ spectrum, albeit with large uncertainties, for the second source. Their position appear to be associated to the galaxy NGC~3341. At its redshift, their 0.5-7 keV luminosities would be $\simeq8\times10^{39}$ erg s$^{-1}$ and $\simeq3\times10^{39}$ erg s$^{-1}$, respectively, thus being possible candidates for Ultra-luminous X-ray sources (ULXs).

\section{Discussion}

The multi-wavelength observations of the triple merging galaxy NGC~3341 presented in this paper gave new insights into the nature of this very peculiar system. The clearest result is on the nature of NGC~3341B: both the X-ray and the optical \textit{SDSS} spectrum agree on a classification as a Compton-thin Seyfert 2 galaxy, confirming with even strongest evidence the claim by B08 based on their \textit{Keck} optical spectrum. All the observed parameters (X-ray and radio luminosities, the $L_x/L_{\mathrm{[{O\,\textsc{iii}}]}}$ ratio, the Eddington rate) are consistent with common values found in local Seyfert galaxies. The only crucial peculiarity of NGC~3341B is, of course, the fact that it does not reside in the center of the main galaxy, but in the nucleus of a dwarf companion. We will extensively discuss this issue later.

The nature of the nucleus of NGC~3341 is less clear. The optical spectrum presented by B08 points towards a classification as a star-forming galaxy, although the $\mathrm{[{N\,\textsc{ii}}]}$ line is somewhat more intense than what generally found in this class of objects, allowing for a possible classification as a `composite' object (Fig.~\ref{3341B_bpt}). It is very interesting to note that B08 report that the optical line profiles are double-peaked, with a velocity separation of 310 km s$^{-1}$. The X-ray spectrum is rather poor, but is consistent with the emission from a starburst: it is very soft, and the total X-ray luminosity is a few times $10^{39}$ erg s$^{-1}$. Adopting the relations presented by \citet{ranalli03}, this luminosity corresponds to a star-formation rate of $\simeq1$ M$_\odot$ yr$^{-1}$, which should be considered only relative to the inner region of the galaxy, due to the high angular resolution of \textit{Chandra}. Indeed, a somewhat larger star-formation rate ($\simeq6$ M$_\odot$ yr$^{-1}$) can be derived from the observed radio luminosity \citep[e.g.][]{bell03,murphy11}, which encompasses all the galaxy.

Interestingly, most of the \textit{Wide-field Infrared Survey Explorer} \citep[\textit{WISE}:][]{wise} mid-infrared emission at 12\micron\, (FWHM$\simeq10$ arcsec) is clearly associated with the nucleus, not with the AGN in NGC~3341B. Indeed, the \textit{WISE} luminosity at 12\micron\, is $1.4\times10^{43}$ erg s$^{-1}$, an order of magnitude larger than what expected from an AGN with the X-ray luminosity of B \citep{gandhi09}. On the other hand, the WISE luminosity is consistent with a star-forming rate of $\simeq4$ M$_\odot$ yr$^{-1}$  \citep{donoso12}, in good agreement with the above estimates derived from the radio and X-ray luminosities. This SFR is slightly below the main sequence of local star-forming galaxies \citep{elbaz11}, given the host stellar mass of the primary galaxy ($\simeq1\times10^{11}$ M$_\odot$: B08), and therefore it appears not to be enhanced by the merger.

Our milliarcsecond resolution radio images obtained with the \textit{EVN} at 1.7 and 5.0 GHz are also more easily reconciled with a starburst dominated scenario. In fact, if the bulk of the radio emission detected with the \textit{EVLA} is due to synchrotron non-thermal emission from relativistic electrons accelerated in supernovae and/or supernova remnants, which have diffused out their acceleration sites, one expects that at the distance of NGC~3341 this radio emission is resolved out by the \textit{EVN}, as observed.

\begin{figure}
\begin{center}
\epsfig{file=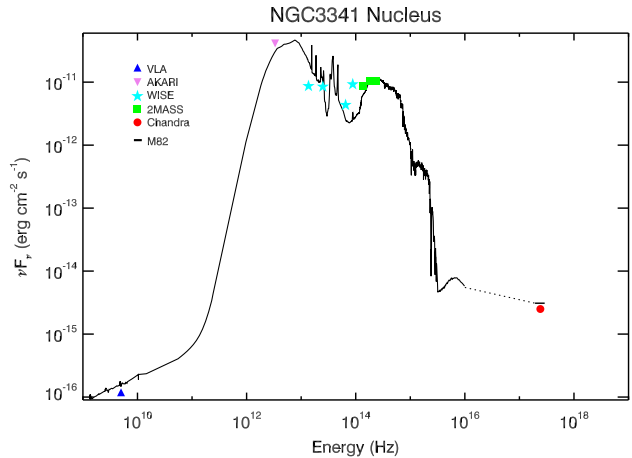,width=\columnwidth}
\end{center}
\caption{\label{seds}Spectral Energy Distribution for the nucleus of NGC~3341. Fluxes are from \textit{VLA} (blue triangle), \textit{AKARI} (violet upside down triangle), \textit{WISE} (cyan star), \textit{2MASS} (green squares), and \textit{Chandra} (red circle). As discussed in the text, we associate all the \textit{WISE} flux to the nucleus. The same choice was adopted for the \textit{AKARI} flux \citep[only the flux at 90\micron\, is reported as secure for this source:][]{akari}. As a comparison, the re-normalized SED of the starburst galaxy, M82, is superimposed on the data. The SED is adapted from \citet{polletta07}, using NED data in the radio band. The X-ray flux of M82 is taken from the nuclear region (1 kpc radius, comparable to the 2 arcsec used for the nucleus of NGC~3341) reported in \citet{strick04}.}
\end{figure}

Therefore, all the multiwavelength data collected in this paper are consistent with the nucleus being a star-forming region (see the Spectral Energy Distribution presented in Fig.~\ref{seds}). Although also the soft X-ray emission can well be due entirely by star-formation, it is still possible that an obscured or intrinsically weak AGN may be present in the nucleus. Low-Luminosity AGN with X-ray luminosities as low as few $10^{39}$ erg s$^{-1}$, or even less, are not uncommon in the local Universe \citep[see e.g.][and references therein]{panessa06,ghosh08,grier11}. Even if the X-ray data do not exclude this possibility, it is also true that we have no clear evidence for activity of the nucleus, apart from the the weak hint from the $\mathrm{[{N\,\textsc{ii}}]}$ emission line. A more powerful AGN can be still be present in the nucleus, if it were obscured by Compton-thick gas. In this case, the intrinsic luminosity of the AGN could be of the order of few $10^{41}$ erg s$^{-1}$ \citep[see e.g.][]{mar12}. However, in this case the AGN contribution to the optical spectrum would be significantly larger, at odds with the (almost) complete lack of signatures in the observed optical spectrum. A possible solution to this problem can be recovered if the AGN is completely enshrouded in a Compton-thick medium, which prevent the ionizing continuum to illuminate in all directions, and allowing even for larger intrinsic X-ray luminosities, because the Compton reflecting medium itself would be absorbed \citep[see e.g.][]{armus07,ueda07,nog09}. The presence of such a `buried' AGN can be unveiled through infrared L-band spectroscopy \citep[e.g.][]{ima06}, or future hard X-rays observations with a focusing telescope with good spatial resolution.

Finally, for NGC~3341C the observational evidence we have collected so far is even scarcer. As noted by B08, the poor optical spectrum of this source presents a $\mathrm{[{N\,\textsc{ii}}]}$ ratio which is inconsistent with a star-forming galaxy, suggesting a classification as a LINER. The very low upper limit on the X-ray luminosity ($3\times10^{38}$ erg s$^{-1}$) is still compatible with this interpretation, but it is questionable that LINERs with such a low luminosity really harbour an AGN at their centre \citep[see e.g.][]{eracl02,flohic06,gonz09}. As in the case of the nucleus of NGC~3341, our data cannot exclude the presence of an AGN with larger intrinsic luminosity, if buried in a $\simeq4\pi$ Compton-thick absorber.

The information we have gathered on the three nuclei of NGC~3341 allows us to depict three different scenarios for this complex system.

\subsection{Scenario 1: AGN multiplet}

AGN triplets are extremely rare, with only a handful reported so far \citep[e.g.][]{djorg07,alonso08,liu11,schaw11}. Such systems are expected to form via a subsequent merger of a dual AGN system which has not coalesced yet \citep[e.g.][]{valt96}, but they should be much rarer than binary mergers \citep{darg11}. For this reason, alternative explanations have been proposed in some AGN triplet candidates, in terms of in situ growth from BH seeds \citep{schaw11}.

Although triple SMBH systems are extremely promising for our understanding of merging and galaxy formation, this appears as the less likely scenario for NGC~3341. Indeed, an accreting SMBH is certainly present in NGC~3341B, which is a common Seyfert 2 galaxy. On the other hand, we have not found any definite evidence of accretion onto the primary nucleus and NGC~3341C. Even if it is not possible to completely exclude accretion activity in either of the two sources, it is very unlikely that both are AGN.

Once assessed, without any doubt, the presence of an AGN in one of the nuclei of NGC~3341, this triple merging system naturally represents an ideal candidate for a dual AGN system. The absence of clear signatures in the optical spectra of the nucleus and NGC~3341C is not a very strong evidence against their AGN activity. However, as summarized in the previous section, no evidence for AGN activity comes out from the radio and X-ray data, either. Therefore, although it is difficult to exclude completely the possibility that at least one of the other two nuclei hosts an AGN, the possibility that NGC~3341 hosts a dual AGN is unlikely at the moment.

In any case, it is interesting to mention that the narrow-line profiles at the nucleus are double-peaked, with a peak-to-peak separation of $\simeq300$ km s$^{-1}$ (B08). Double-peaked AGN are widely studied, because considered good candidates to host dual AGN, although with a very low overall efficiency \citep[only a few percent of double-peaked AGN are indeed dual AGN, most of them arising instead from other mechanisms, like bipolar outflows:][]{fu12}. Notwithstanding, simulations do show that, even if the double peaks comes from the same NLR, its kinematics may still be influenced by the relative motions of the two SMBHs \citep[see][but the simulations are valid only for major mergers]{blecha13}. In the case of the nucleus of NGC~3341, the double-peaked spectrum does not belong to an AGN, but to a star-forming galaxy. Double-peaked galaxies hosting starbursts are yet to be studied in detail as done for AGN. A systematic study performed by \citet{pil12} suggests that two $\mathrm{[{H\,\textsc{ii}}]}$ regions in separate regions of the galactic disc can generate the double-peaked emission lines. Given the slit width adopted by B08 (0.75\arcsec), these star-forming regions cannot be separately resolved in the \textit{Chandra} image presented in Fig.~\ref{3341_softX}. 

\subsection{Scenario 2: Offset AGN}

Major mergers are believed to impress kick velocities up to several thousand km s$^{-1}$ on the SMBH generated by the coalescence of the original two SMBHs, as a result of the Gravitational Waves emission during the process \citep[see e.g.][for a review]{kom12}. The new SMBH will then recoil and start oscillating about the core of the host galaxy. Interestingly, the intrusion of a third galaxy in the merging system may favour the displacement of one or more SMBHs in the mechanism know as gravitational slingshot \citep{saslaw74,hl07}. A displacement of $\simeq5$ kpc from the nucleus, as observed in the case of NGC3341, is in agreement with the predictions from hydrodynamic/N-body simulations, considering the low accretion rate of NGC~3341B \citep[higher accretion rates are possible only during pericentric passages for short-period recoiling oscillations or before leaving the central dense region for long-period ones:][]{blecha11}. In any case, the corresponding recoiling velocity cannot be directly measured in NGC~3341B as in other candidates because this source is an optical Type 2 object, and therefore no velocity displacement can be measured in the (unobservable) broad lines.

Recoiling AGN can carry with them bound matter within a region roughly on the scale of the BLR \citep[e.g.][]{kom12}, thus well within the sublimation radius. Therefore, recoiling AGN can, in principle, be X-ray obscured as NGC~3341B, if the X-ray absorber is located at the distance of the BLR \citep[as now commonly found in local Seyfert galaxies: e.g.][and references therein]{ris05,bianchi09c,bmr12}, but it should not be an optical Type 2 object, because any dusty absorber outside the BLR itself is not bound to the recoiling AGN any more \citep[e.g.][]{civ12}. If, instead, the BLR reddening is due to a larger scale material, like a galactic dust lane, it is still possible that a type 2 object, like NGC~3341B, is a recoiling AGN. 

However, the offset AGN scenario is strongly disfavoured by the morphology of the system, which, as well as being a minor (and not major as generally assumed for recoiling AGN) merger, is clearly constituted by three still separated galaxies, each with its own SMBH at the center. The offset AGN hypothesis would therefore require the unlikely occurrence that the SMBH at the center of the primary galaxy has been kicked at the center of NGC~3341B. A \textit{Hubble Space Telescope} observation could help in understanding in detail the morphology of the merging system, and rejecting conclusively the recoiling scenario.

Finally, we would like to stress that, differently from other hyper-luminous (L$_x$ $>10^{41}$ erg s$^{-1}$) off-nuclear X-ray sources \citep[e.g.][and references therein]{farr09,jonker10}, NGC~3341B is an optically confirmed AGN and the nucleus of a dwarf galaxy in a merging system. Therefore, alternative suggestions proposed for this class of sources (like an Intermediate Mass BH or a very bright supernova) do not apply in this case.

\subsection{Scenario 3: AGN triggering by the merging process}

It is possible that the AGN activity in NGC~3341B is completely unrelated to the ongoing merging process between the three galaxies. If the fuelling of Seyfert-like objects is mainly powered by random accretion events \citep[e.g.][]{kp07}, it is possible that we are simply witnessing a galaxy hosting an AGN which is serendipitously interacting with other galaxies, without implying that the interaction itself is the origin of its nuclear activity. As already noted by B08, although the AGN fraction in low-mass galaxies is very low, there are other examples of AGN hosted in galaxies with luminosities similar to B, but they are always isolated systems. On the other hand, galaxy mergers and interactions are believed to be the critical players in the supply of interstellar gas up to the innermost region of the galaxies, allowing for the ultimate fuelling of the AGN activity \citep[e.g.][]{ellison11}. It is therefore relevant to speculate on the relation between the AGN activity in B and the complex merging system in which it resides.

In the hierarchical growth of structures, minor mergers are much more common than major mergers, and therefore it is crucial to understand their role in AGN triggering and galaxy evolution. In this system, it appears that only one AGN is present, with the exceptional peculiarity that it is the only known well-studied minor merger in which the AGN is located in at least one secondary galaxy instead of the primary. Indeed, such occurrence is basically not observed in large systematic studies of galaxies in minor pairs \citep{wg07}. This can be explained by an obvious observational bias, since it is extremely difficult to identify and separate a smaller and dimmer satellite galaxy from its primary companion. In this sense, X-ray observations, as in the case of NGC~3341, can be extremely powerful in finding AGN activity in the secondary galaxy, but, due to the limited X-ray spatial resolution and throughput, they become rapidly ineffective outside of the local Universe.

Unfortunately, most of the theoretical studies performed so far have been focused on the induced star formation and nuclear activity in major mergers and, therefore, there is a lack of detailed time- and spatially-resolved predictions on the evolution of NGC 3341-like systems. The numerical simulations of \cite{cox08} represent one of the best effort to address the issue of starburst episodes (that can be broadly assimilated to AGN activity) in minor mergers. They found that in case of large galaxy mass ratios, the SFR of the primary galaxy remains basically unchanged during the interaction, while the satellite galaxy may undergo a significant enhancement of its SFR as result of a larger tidal force effect. This could lead to a faster timescale for the AGN triggering in the secondary nucleus with respect to the primary galaxy, thus producing short-lived (and therefore rare) systems like NGC~3341. According to \citet{cox08}, for a merger mass ratio of 1:23.8 (i.e. similar to NGC~3341 B:A, B08)  very little star formation (1-3 M$_\odot$ yr$^{-1}$) is expected in the primary galaxy, that is consistent with our estimates.

However, the efficiency of this process strongly depends on the real parameters of the system. For example, the fraction of SMBH presence in satellite dwarf galaxies is believed to be well below unity \citep{vol08}: without the BH, of course no AGN activity can be triggered. Moreover, the gas fraction in the secondary galaxy can be very low, because it can be easily stripped in the merging process, thus quenching the onset of AGN activity for lack of fuel \citep{cox08}.

\section{Conclusions}

We have presented a comprehensive, panchromatic analysis of the triple-merging system in NGC~3341. We confirm and strengthen the classification of NGC~3341B as an AGN, with typical parameters of local Seyfert 2 galaxies. On the other hand, the nucleus of NGC~3341 is compatible with a star-forming region, while NGC~3341C, undetected both in X-rays and in radio, still cannot be definitely classified, but it is very unlikely that it could host an AGN.

All the information gathered seems to point to the intriguing scenario of an AGN triggered in a satellite galaxy of a minor-merging system. The rarity of such a system can be ascribed to a trivial observational bias, which makes it difficult to separate an intrinsically weak activity, due to the low mass of the satellite, from the nucleus of the close primary galaxy. However, the onset of AGN activity in the secondary galaxy can be intrinsically rare or characterized by a very short time-scale. Further observational and theoretical efforts are needed in order to fully understand the role of minor mergers in the hierarchical growth of structures and AGN triggering.

The triple merging system in NGC~3341 still deserves a deeper investigation. On one hand, infrared L-band spectroscopy is needed to find any evidence for AGN activity in the nucleus of the primary galaxy, or confirm with even higher confidence its absence. On the other hand, it would be extremely interesting to look for any evidence of shocked gas and perform a detailed study of the gas morphology of this unique system, by mapping the cold and warm molecular gas through CO and H$_2$ emission, as well as more ionised gas through $\mathrm{[{O\,\textsc{iii}}]}$ and H$\alpha$ emission.

\section*{Acknowledgements}

We thank the anonymous referee for helping us in improving the paper. We also thank M. Chiaberge, E. Giallongo, M. Guainazzi, A. Lamastra, and N. Menci for useful discussions. SB, FLF and GM acknowledge financial support from PRIN2011 (grant 2010NHBSB). MAPT acknowledges support by the Spanish MINECO through grant AYA 2012-38491-C02-02, cofunded with FEDER funds. 

Funding for SDSS-III has been provided by the Alfred P. Sloan Foundation, the Participating Institutions, the National Science Foundation, and the U.S. Department of Energy Office of Science. The SDSS-III web site is http://www.sdss3.org/. SDSS-III is managed by the Astrophysical Research Consortium for the Participating Institutions of the SDSS-III Collaboration including the University of Arizona, the Brazilian Participation Group, Brookhaven National Laboratory, University of Cambridge, Carnegie Mellon University, University of Florida, the French Participation Group, the German Participation Group, Harvard University, the Instituto de Astrofisica de Canarias, the Michigan State/Notre Dame/JINA Participation Group, Johns Hopkins University, Lawrence Berkeley National Laboratory, Max Planck Institute for Astrophysics, Max Planck Institute for Extraterrestrial Physics, New Mexico State University, New York University, Ohio State University, Pennsylvania State University, University of Portsmouth, Princeton University, the Spanish Participation Group, University of Tokyo, University of Utah, Vanderbilt University, University of Virginia, University of Washington, and Yale University. 

The European VLBI Network is a joint facility of European, Chinese, South African and other radio astronomy institutes funded by their national research councils. e-VLBI research infrastructure in Europe is supported by the European Union's Seventh Framework Programme (FP7/2007-2013) under grant agreement number RI-261525 NEXPReS.

This publication makes use of data products from the Wide-field Infrared Survey Explorer, which is a joint project of the University of California, Los Angeles, and the Jet Propulsion Laboratory/California Institute of Technology, funded by the National Aeronautics and Space Administration. 

This research is based on observations with AKARI, a JAXA project with the participation of ESA.

\bibliographystyle{mn2e}
\bibliography{sbs}

\label{lastpage}

\end{document}